\documentclass[aps,prb, twocolumn, superscriptaddress, longbibliography, nobalancelastpage]{revtex4-1}

\usepackage{times}
\usepackage{amsmath}
\usepackage{amssymb}
\usepackage{graphicx}
\usepackage[caption=false, position=top, singlelinecheck=off, justification=raggedright]{subfig}
\usepackage{color}
\usepackage{pst-node}
\usepackage[unicode]{hyperref}
\definecolor{ACSblue}{RGB}{13,84,166}
\hypersetup{unicode=true, colorlinks=true, linkcolor=black, citecolor=black,urlcolor=black }
\usepackage{float}
\usepackage[normalem]{ulem}
\usepackage{multirow}
\usepackage{hhline}

\usepackage{mathrsfs,braket}
\usepackage{amssymb, amsbsy, amsmath, latexsym, dsfont, array, layout, graphicx, mathrsfs, color, ulem, bm}

\usepackage{changepage}  

\usepackage{xcolor}
\usepackage{tikz}
\definecolor{ACSblue}{RGB}{38,51,128}

\newcommand{\ACsquare}{
  \tikz[baseline=0ex]{
    \fill[ACSblue] (0,0) rectangle (0.8em,0.8em);
  }
}

\usepackage[T1]{fontenc}
\usepackage[scaled=1.0]{sourcesanspro}

\newcommand{\ACSFont}[1]{
  {
    \fontfamily{SourceSansPro-TLF}\fontseries{sb}\selectfont 
    \large 
    \textcolor[RGB]{38,51,128}{#1}%
  }
}

\newcommand{\ACSection}[1]{
  \noindent
  \hspace*{-0.4em} \ACsquare\hspace{0.01em} 
  \ACSFont{#1}\par
  \vspace{0.3\baselineskip}
}

\newcommand{\ACSSISubtitle}[1]{
  {
\fontfamily{SourceSansPro-TLF}\fontseries{sb}\selectfont
   \normalsize #1}
}
 

\makeatletter
\renewcommand*{\fnum@figure}{\textbf{Figure \thefigure}}

\renewcommand*{\@caption@fignum@sep}{\textbf{.} }
\makeatother

\definecolor{orcidlogocol}{HTML}{A6CE39}

\newcommand{\orcidicon}{
  \tikz[baseline=-0.6ex]{
    \node[draw=orcidlogocol, fill=orcidlogocol,
          circle, inner sep=0pt, minimum size=1.6ex]
         {\scriptsize\textcolor{white}{iD}};
  }
}

\begin{document}

\title{Geometric Spin-Orbit Coupling Resolves the Contradictory CISS Effect in Chiral Single Molecules}

\author{Shu-Zheng Zhou}
 \affiliation{School of Physics and Wuhan National High Magnetic Field Center,
Huazhong University of Science and Technology, Wuhan 430074, People's Republic of China.}

\author{Xi Sun}
 \affiliation{School of Physics and Wuhan National High Magnetic Field Center,
Huazhong University of Science and Technology, Wuhan 430074, People's Republic of China.}

\author{Kai-Yuan Zhang}
 \affiliation{School of Physics and Wuhan National High Magnetic Field Center,
Huazhong University of Science and Technology, Wuhan 430074, People's Republic of China.}

\author{Hua-Hua Fu}
\altaffiliation{Corresponding author.\\ hhfu@hust.edu.cn}
\affiliation{School of Physics and Wuhan National High Magnetic Field Center,
Huazhong University of Science and Technology, Wuhan 430074, People's Republic of China.}
\affiliation{Institute for Quantum Science and Engineering, Huazhong University of Science and Technology, Wuhan, Hubei 430074, China.}

\date{\today}

\begin{abstract}
\noindent
\noindent{\bfseries ABSTRACT:} Some studies have reported clear chirality-induced spin selectivity (CISS) effect in four classes of chiral single molecules with remarkable spin polarization. In contrast, a recent high-precision measurement involving nearly a thousand individual tests failed to detect significant CISS signals in the same molecular systems (J. Am. Chem. Soc. 2025, \textbf{147}, 25043). These conflicting results cast doubt on whether CISS truly occurs in these chiral systems at the single-molecular level. To resolve this discrepancy, we develop a theoretical framework incorporating geometric spin-orbit coupling and environmental decoherence, enabling systematic study of the CISS in four chiral single molecules with distinct geometries and sizes. Our calculations show that the CISS effect is completely suppressed in both strong-coherence and strong-decoherence regimes, but becomes pronounced in the intermediate-decoherence regime, where observable spin polarization emerges. In the strong-coherence regime, both electron-electron interaction and electron-vibration coupling enhance the CISS effect: the former is more effective in large molecules, whereas the latter plays a more significant role in smaller ones. Increasing temperature further enhances spin polarization. The proposed mechanism unifies contradictory experimental observations and reveals how the CISS effect evolves from regular helical (helical symmetric) to irregular helical (point-symmetric or axially symmetric) chirality. This framework thus provides a basis for unifying CISS phenomena across single-molecule systems, regardless of their specific molecular configurations or symmetry classes.  

\end{abstract}
\maketitle

\ACSection{INTRODUCTION}
\noindent The chirality-induced spin selectivity (CISS) effect refers to the ability of chiral materials to filter electron spin in the absence of an external magnetic field. Unlike conventional spin polarization mechanisms that rely on magnetic material, CISS effect may appear in nonmagnetic materials while exhibit high spin polarization (SP), making it potential for spintronic devices that avoid magnetic components. This unique SP mechanism has since been explored in fields ranging from physics to electrochemistry and energy science \cite{1,2,3,4,5}. The CISS effect was first observed in photoinduced electron transport through double-stranded DNA \cite{6,7} and has since been validated in various other chiral systems, including chiral crystals \cite{8,9}, helical superstructures \cite{10,11,12,13}, chiral perovskites \cite{14,15,16,17}, and chiral supramolecular assemblies \cite{18,19}. These findings have led to a general consensus that the CISS is ubiquitous in chiral materials. However, a significant contradiction has emerged regarding the experimental evidence for this effect in chiral single molecules (CSMs). Some studies have reported clear CISS phenomena in four different CSM families \cite{20,21,22,23,24,25,26,27,28}, including atropisomers (1S/1R), enantiomers (2S/2R), bowl-shaped Subphthalocyanine derivatives, and dipolar L-histidine, with SP reaching as high as 40\%. In contract. a recent high-precision measurement involving nearly a thousand tests on the same CSMs failed to detect distinguishable CISS signals \cite{29}. These conflicting results pose a serious challenge to the universality of CISS effect at the single-molecule level. 

To resolve this experimental discrepancy, a unified theoretical framework is urgently needed to understand the CISS effect in CSMs. Notably, in their review on electron transport through single molecules, F. Evers and colleagues identified the theory of CISS in single-moelcule systems as one of the three long-standing unsolved problems in the field \cite{29_1}. This assessment makes clear that developing a reliable theoretical model is not only essential for clarifying the current controversy but also a necessary step to advance single-molecule science. Experiments have demonstrated that, under identical measurement conditions, the geometric configuration of CSMs significantly influences the observed results \cite{24,25,26,27,28}. This suggests that the origin of CISS lies inherently in the molecular structure itself, with spin-orbit coupling (SOC) remaining one of the most plausible key mechanisms. However, in the theoretical framework of CISS based on the SOC, the origin of SOC itself remains a topic of ongoing debate \cite{30,31,32,33,34,35,36}. Early theoretical explanations of CISS were built upon atomic intrinsic SOC (Figure 1\textbf{a}). Although this mechanism provides an intuitive picture for understanding spin-polarized conducting electrons \cite{37,38,39,40}, the intrinsic SOC in light atoms of organic molecules is exceedingly weak, failing to account for the high SP observed in purely organic chiral molecules at room temperature. When experiments fail to observe CISS, attributing this absence solely to weak intrinsic SOC readily yields low SP, consistent with such null results, but this reasoning often overlooks other critical factors that can suppress the CISS effect, such as electron decoherence \cite{29}. To overcome this limitation, some researchers have proposed an alternative SOC model driven by an effective electric field, where the SOC strength is attributed to an electrostatic potential gradient along the helix axis (Figure 1\textbf{b}) \cite{26,41,42}. Although this mechanism can formally enhance the effective SOC strength, the magnitude of potential gradient required to achieve for such high SP often exceeds the practical tolerance range of CSMs \cite{34,42}. Moreover, accurately defining such a gradient in geometrically irregular single-molecule systems remain challenging \cite{43,44,45}.

\begin{figure*}[t]
\includegraphics[width=1.6\columnwidth]{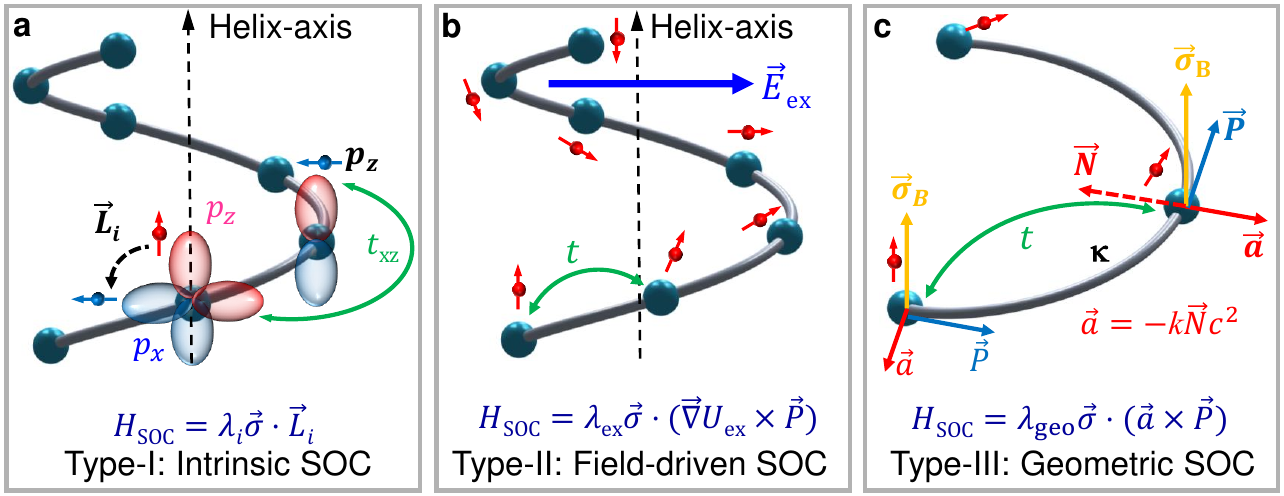}
\caption{Three SOC mechanisms that can be used to explain the generation of spin polarization. (a) Type-I: Atomic intrinsic SOC between different $P$-orbital components (e.g., $P_x$, $P_y$, $P_z$). (b) Type-II: Effective field-driven SOC from the electrons moving along the chiral paths, equivalent to an external electric field $\vec{E}_{ex}$=-$\nabla U$ applied to the system. (c) Type-III: Geometric SOC due to the relativistic effect for electrons moving on curved paths.}
\label{fig1}
\end{figure*}

On the other side, existing CISS theories predominantly rely on simplified one-dimensional (1D) regular helical chain models \cite{19,45_01}. Although such models provide instructive physical pictures within traditional SOC framework and can explain certain typical experimental phenomena, for example, that the accumulated SOC with increasing helix length enhances spin-filtering efficiency, which explains why longer DNA strands typically exhibit higher SP than shorter ones \cite{19}. However, they also introduce another persistent and unresolved issue, that is, when a helix is so short that it does not complete even one full turn, or is merely a short helical arc, the helical potential within these traditional frameworks is greatly suppressed. Under such conditions, does the CISS effect still exist? if so, how does the CISS effect evolve when a non-periodic curved geometry further deviates from regularity? This challenge reveals a limitation of traditional SOC-based frameworks when applied to finite-length, non-periodic curved conduction pathways. Compounding this issue, advances in the synthesis and fabrication of CSMs have produced systems with various symmetry types of chirality: point-, axial-, and planar-symmetry chirality, many of which exhibit high SP despite lacking helical symmetry \cite{45_1,45_2,45_3,45_4}. These non-helical chiral materials are increasingly difficult to reconcile with helix-based SOC models. Thus, an urgent need remains for a CISS theory capable of describing the entire continuum from regular helices to highly irregular or transitional helical structures. Addressing this gap is necessary to build a unified CISS theory for CSMs that can account for the conflicting observations across the four molecular classes. 

\begin{figure*}[t]
\includegraphics[width=1.50\columnwidth]{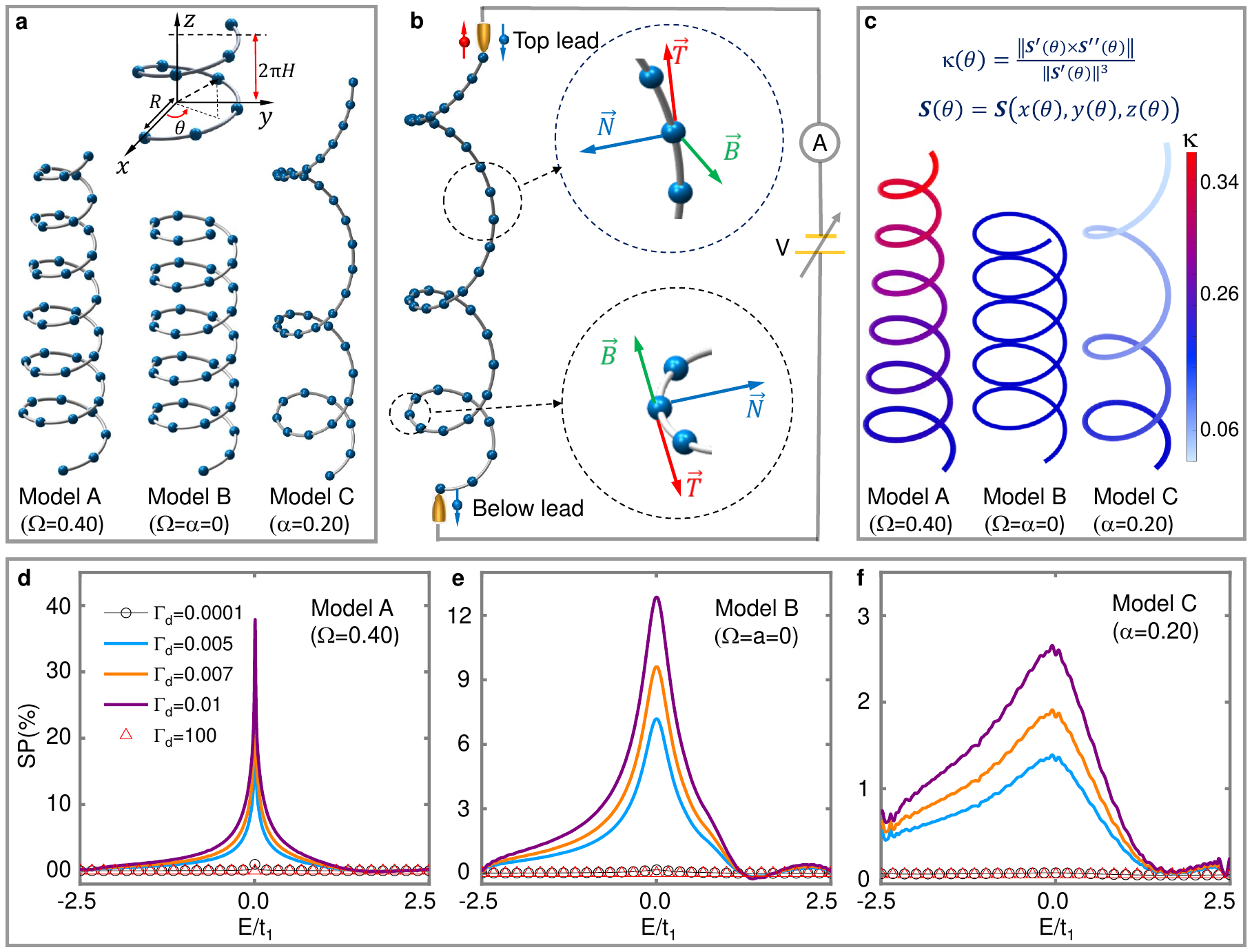}
\caption{Spin polarization for three chiral chain geometries along with the transport device configuration, local coordinate vectors and spatial curvature. (a) Three representative single-chiral chain geometries: cone-shaped (Model A), uniform (Model B), and progressive-pitch (Model C). (b) Transport device based on Model C, coupled to two nonmagnetic electrodes. The local tangent ($\vec{T}_i$), normal ($\vec{N}_i$) and binormal ($\vec{B}_i$) vectors are shown in red, green and magenta, respectively. (c) Spatial variation of the curvature $\kappa$ in the three helix models. \textbf{d}-\textbf{f} Spin polarization versus energy $E$ for Model A ($\Omega$=0.4), Model B ($\Omega$=$\alpha$=0), and Model C ($\alpha$=0.20).}
\label{fig2}
\end{figure*}

Therefore, to investigate the generation conditions and evolution trends of the CISS effect during the degradation of regular helical chains into aperiodic curved arcs or irregular helicoids, it is necessary to go beyond conventional SOC mechanisms and seek a new one that can accommodate different chiral molecular geometries. Surveying all past CISS experiments on CSMs, a common features: all CISS effect is based on conduction electrons or photoexcited electrons that traverse a twisted pathway within the chiral molecules. When an electron is confined to move along a curved or twisted path, its direction changes continuously, and this geometric variation impacts an effect on the electron spin, generating a new type of effective SOC, referred to as geometric SOC \cite{45,46,47,48}. This kind of SOC can exceed the intrinsic SOC in organic molecules by one to two orders of magnitude \cite{45}. Moreover, a closely related effect, known as curvature-induced SOC, has already been experimentally observed in carbon nanotubes \cite{48_1}, which can be understood as a manifestation of geometric SOC in curved quasi-one-dimensional systems. These features of geometric SOC provide strong support for the high SP observed experimentally. In addition, because this SOC is directly linked to the specific conduction pathways within chiral molecules, it addresses precisely the deficiency highlighted in a recently study by Naaman and colleagues, namely, that all previous CISS theories neglected the explicit role of the electron's transport pathway \cite{48_2}. Thus, geometric SOC offers a fresh perspective on the physical origin of the CISS effect in CSMs. Based on this mechanism, we have established a theoretical framework for CISS in geometrically tunable helices, systematically revealing the evolution of SP through key conformational deformations, from cone-shaped to uniform then to progressive-pitch helical structures. By incorporating geometric SOC together with environmental decoherence, we have constructed an effective model applicable to CSMs of different geometries, successfully reconciling previously contradictory experimental observations \cite{29}.\\

\ACSection{RESULTS AND DISCUSSION}
\noindent {\fontfamily{SourceSansPro-TLF}\fontseries{b}\selectfont Geometric SOC and related theoretical Models.}
To go beyond conventional 1D uniform helical chains (Figure 2\textbf{a}, Model B), we introduce two non-uniform helices: a cone-shaped chain with radius decreasing along the helical axis, forming a 3D tapered profile (Figure 2\textbf{a}, Model A); and a progressive-pitch chain with pitch increasing along the axis (Figure 2\textbf{a}, Model C). By tuning the radius parameter $r$ (Model A) or pitch parameter $z$ (Model C), both can be continuously transformed into the uniform helix (Model B), enabling a unified Hamiltonian description of their CISS phenomena. To establish an effective tight-binding model, we discretize each chain into a lattice model by dividing the total arc length into $N$ equal segments, yielding $N$+1 lattice sites. The spatial coordinates of the $i$th site are then given by 
\begin{equation}
\textbf{S}_i=\textbf{S}_i(\theta_i)=[r\mathrm{cos}{\theta_i}, \zeta r\mathrm{sin}{\theta_i}, H{\theta_i}+\alpha\theta^2_i]^T.
\end{equation}
Note that for a cone-shaped helix (Model A), the radius decreases linearly as $r(\theta_i)=R(1-\frac{\Omega}{2\pi Q}\theta_i)$, whereas for the uniform helix (Model B) it remains constant, $r(\theta_i)=R$. For the progressive-pitch helix (Model C), the axial coordinate expands according to $z(\theta_i)=H\theta_i+\alpha\theta^2_i$, where $\theta_i=i\Delta\theta$ denotes the angular coordinate of lattice site $i$=(1, 2,$\cdot\cdot\cdot$, $N$+1). The parameter $\zeta$ indicates molecular chirality (left-handed for $\zeta$=-1 and right-handed for $\zeta$=1), $Q$ is the total number of turns, and $\theta$$\in$[0, 2$\pi$\textit{Q}]. To characterize irregular helices, we introduce two deformation parameters: a tapering parameter $\Omega$, which controls the sharpness of the conical profile in Model A (larger $\Omega$ corresponding to a sharper tip), and an unwinding parameter $\alpha$, which governs the rate of pitch expansion in Model C (larger $\alpha$ leading to faster uncoiling). For $\Omega$=0 and $\alpha$=0, the uniform helix (Model B) is recovered.

When electrons move along curved paths in CSMs, the geometric SOC is introduced due to the effective acceleration emerging in a non-inertial frame (details in Supporting Information, Sections I-A). The strength of this SOC is determined by the molecular geometric curvature and the accumulated Berry phase, linking it to molecular spatial configuration rather than atomic number. Geometric SOC is completely different from traditional Rashba and Dresselhaus SOC (Supporting Information, Sections I-B), and its strength can be two orders of magnitude higher than the atomic intrinsic SOC (Supporting Information, Sections I-C and I-D), providing a key basis for high SP in CISS and making it particularly suitable for explaining CISS phenomena in light-element chiral molecules. Within the Frenet-Serret framework \cite{46,47}, the tangent and normal vectors at site $i$ are denoted by $\vec{T}_i$ and $\vec{N}_i$, respectively (Figure 2\textbf{b}), thus this local effective acceleration for electrons can be expressed as $\vec{a}_{i}$=$-\kappa_{i} c^2 \vec{N}_{i}$ (where $\kappa_{i}$ is the local curvature and $c$ is the speed of light), which subsequently induces the geometric SOC Hamiltonian $\mathcal{H}_{\mathrm{geo}}$=$\hbar (\vec{a}\times\vec{p})\cdot \vec{\sigma}/{(4mc^{2})}$=$-\sum_{i} \hbar \kappa_{i} (\vec{N}_{i}\times\vec{p}_{i}) \cdot \vec{\sigma}/{(4m)}$, where $\vec{p}_{i}$ is the momentum, $m$ is the electron mass, and $\vec{\sigma}$ is the Pauli vector (derivation details given in Supporting Information, Sections I-A).

Given that $\mathcal{H}_{\mathrm{geo}}$ is determined and that each helical chain is coupled to two nonmagnetic electrodes, the full tight-binding Hamiltonian of the SCM-based devices is $\mathcal{H}=\mathcal{H}_{\mathrm{mol}}+\mathcal{H}_{\mathrm{geo}}+\mathcal{H}_{\mathrm{el}}+\mathcal{H}_{\mathrm{d}}$. The molecular part $\mathcal{H}_{\mathrm{mol}}$ can be discretized in a Wannier basis as $\mathcal{H}_\mathrm{mol}=\sum^{N}_{i=1}(\varepsilon_{i}c^{\dag}_{i}c_{i}+U_i n_{i,\uparrow}n_{i,\downarrow})$+$\sum^{N}_{n=1}(t_{1}c^{\dag}_{i}c_{i+1}+t_{2}c^{\dag}_{i}c_{i+2}+\mathrm{H}.\mathrm{c}.)$, where $c^{\dag}_{i}=(c^{\dag}_{i,\uparrow},c^{\dag}_{i,\downarrow})$ is the creation operator at the $i$th site, $\varepsilon_{i}$ is the onsite energy, $U_i$ the onsite Coulomb repulsion, and $t_1$ and $t_2$ represent the nearest-neighbor and next-nearest-neighbor hoppings, respectively. Due to equal arc-length discretization, these hoppings can be treated as constants. Similarly, $\mathcal{H}_\mathrm{geo}$ can be discretized as 
\begin{align*}
\mathcal{H}_\mathrm{geo}=
&i\frac{\tilde\lambda_{\mathrm{1}}}{2}\sum^{N}_{i=1}c^{\dag}_{i}(\kappa_{i+1}+\kappa_{i})(\sigma_{i+1}+\sigma_{i})c_{i+1}\\
&i\frac{\tilde\lambda_{\mathrm{2}}}{2}\sum^{N-1}_{i=1}c^{\dag}_{i}(\kappa_{i+2}+\kappa_{i})(\sigma_{i+2}+\sigma_{i})c_{i+2}+\mathrm{H}.\mathrm{c.}
\end{align*}
where $\tilde\lambda_{\mathrm{1}}$=$\frac{\hbar^2}{16mS}$ and $\tilde\lambda_{\mathrm{2}}$=$\frac{\hbar^2}{32mS}$ with $S$ being the arc length between two adjacent lattices (Supporting Information, Sections I-C). The unit binormal vector $\vec{B}_i$=$\vec{T}_i$$\times$$\vec{N}_i$ defines the spin polarization axis via $\sigma_{i}$=$\vec{\sigma} \cdot \vec{B}_i$. This effective SOC acts as a momentum-dependent magnetic field locked to the local Frenet-Serret frame, directly coupling the electron spin to the geometric chirality of single molecules.

The third term, $\mathcal{H}_{el}$, consists of two parts: $\mathcal{H}_{el}=\mathcal{H}_l+\mathcal{H}_{lc}$. The former $\mathcal{H}_l$=$\sum_{k\beta}\varepsilon_{0}a^{\dag}_{\beta k}a_{\beta k}+\sum_{\beta k}t_{0}(a^{\dag}_{\beta k}a_{\beta k+1}+\mathrm{H}.\mathrm{c.})$ describes the electrons within the top ($\beta=t$) and bottom ($\beta=b$) electrodes. Each electrode is modeled as a semi-infinite metallic chain with onsite energy $\varepsilon_0$ and nearest-neighbor hopping $t_0$. The coupling between the SCM and the electrodes is given by $\mathcal{H}_{lc}=\sum_{\beta}\Gamma_{\beta}a^{\dag}_{\beta{1}}c_{n\beta}+\mathrm{H}.\mathrm{c.}$ Here, $a^{\dag}_{\beta k}$ creates an electron at site $k$ in electrode $\beta$; the SCM couples to the site $n_t$=1 in the top and the site $n_b$=$N$+1 in the bottom electrode (Figure 2\textbf{b}).

CISS experiments in CSMs are usually performed at room temperature, where inelastic scattering from thermal vibrations destroys electron phase memory. To simulate this decoherence, we use the B\"{u}ttiker virtual-electrode technique \cite{49,49_1,49_2}, coupling a virtual electrode to each site of molecule through the parameter $\Gamma_d$. These virtual electrodes carry no net current and only erase phase information. Crucially, $\Gamma_d$ is independent of molecule-electrode coupling and bias voltage. When only thermal fluctuations are considered, the chosen $\Gamma_d$ range [0.1, 0.5]$\times10^{-3}$ eV corresponds to a decoherence temperature of about 200–600 K, covering typical room-temperature conditions. Conversely, at $T$=300 K, the smaller values in this $\Gamma_d$ range yield an electron coherence length on the order of tens of nanometers, consistent with the coherence length ($\approx$ 30 nm) measured in Au electrodes under room-temperature measurements \cite{37}. Therefore, $\Gamma_d$ provides a physically sound model for coherence and decoherence in SCM systems. Using this technique, we may construct the corresponding Hamiltonian as $\mathcal{H}_{d}=\sum_{i,k}\varepsilon_{0}d^{\dag}_{ik}d_{ik}+\sum_{ik}(t_{d}d^{\dag}_{i,k+1}d_{ik}+\mathrm{H}.\mathrm{c.})+\sum_{i}({\Gamma_d}c^{\dag}_{i}d_{i,1}+\mathrm{H}.\mathrm{c.})$, with $\varepsilon_0$, $t_d$ and $d^{\dag}_{ik}$ describing the electron energy, hopping and creation operator in virtual electrodes. Imposing the  condition of zero net current on each virtual electrode and combining the Landauer-B\"{u}ttiker formula \cite{50,51} with Green's function theory \cite{52,53}, the spin-up (spin-down) conductance $G_{\uparrow(\downarrow)}$ is calculated and then SP is obtained as SP=$(G_{\uparrow}$-$G_{\downarrow})/(G_{\uparrow}$+$G_{\downarrow})$ (Supporting Information, Section II). \\ 

\begin{figure*}[t]
\includegraphics[width=1.5\columnwidth]{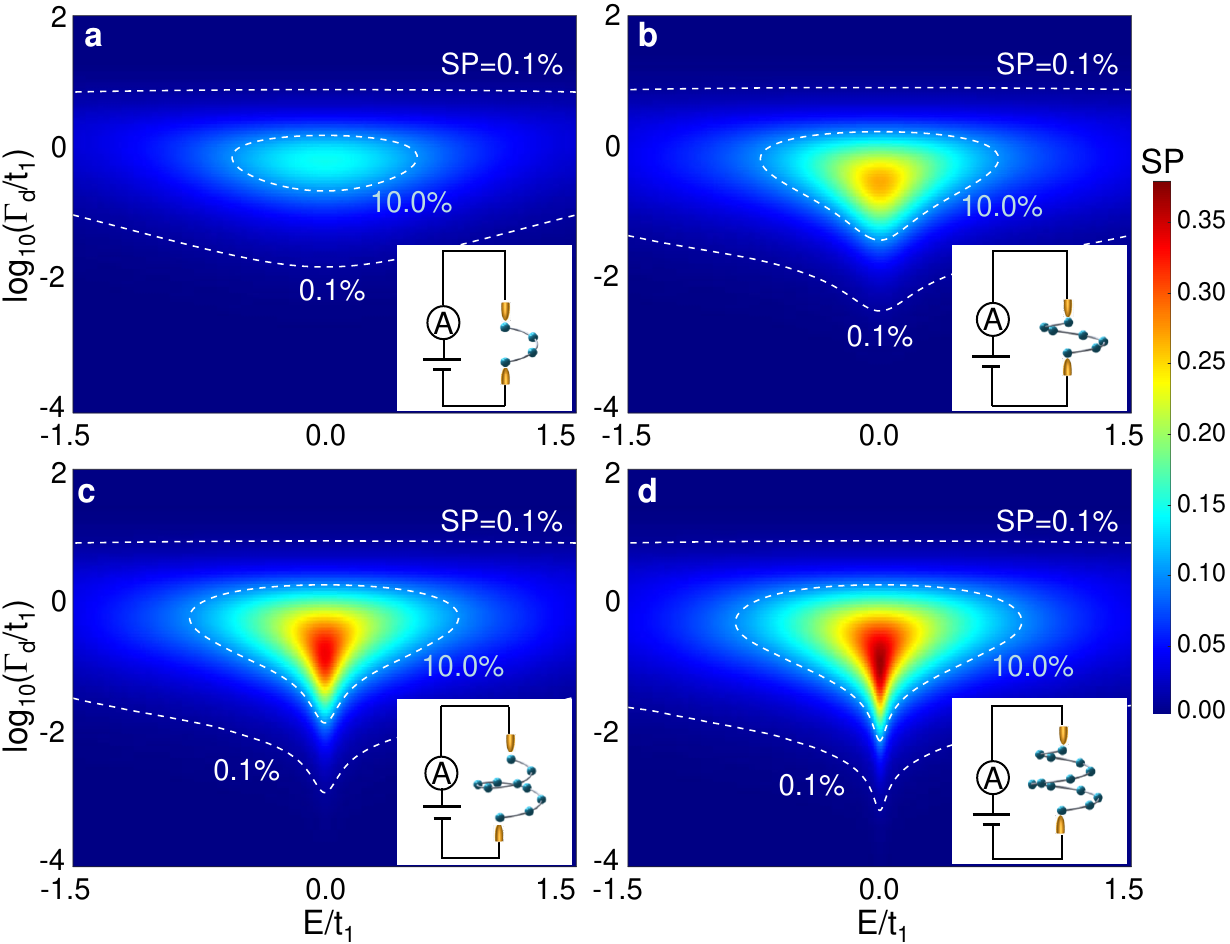}
\caption{Phase diagram of spin polarization versus the decoherence strength log$_{10}$(${\Gamma_{d}/t_1}$) and the energy $E$ for four types of chiral molecular junctions based on a half-turn (a), a full-turn (b), a one-and-a-half-turn (c), and a two-turn helix (d). Three contour lines for SP=1.0\%, 10.0\% and 20.0\% are plotted.}
\label{fig3}
\end{figure*}

\noindent {\fontfamily{SourceSansPro-TLF}\fontseries{b}\selectfont Evolution of CISS Under Structural Deformation.}
Before performing numerical calculations, we evaluated the roles of key structural parameters on CISS, specifically the hopping parameters $t_{1(2)}$ and the SOC parameters $\lambda_{1(2)}$. A canonical transformation applied to $\mathcal{H}_{\mathrm{geo}}$ reveals that the nearest-neighbor SOC does not contribute to SP, whereas the next-nearest-neighbor SOC exhibits significant spin-flip features (Supporting Information, Section III-A), indicating that the former has no contribution on CISS, a conclusion consistent with previous findings for traditional SOC in single-helical chains \cite{37}. Therefore, we adopt the next-nearest-neighbor SOC ($\tilde\lambda_1$=0 while $\tilde\lambda_2\neq$0). Given that the effect of $t_2$ on conductance is much weaker than $t_1$ in CSMs (Supporting Information, Section III-B), we only consider the nearest-neighbor hopping and take $t_1$=0.1 eV based on the typical electronic properties of organic molecules. To highlight the crucial role of geometric SOC in CISS, we first consider the CSM's Hamiltonian in the single-particle framework ($U_i$=0) and then gradually incorporate many-body correlations. 

\begin{figure*}[t]
\includegraphics[width=2.0\columnwidth]{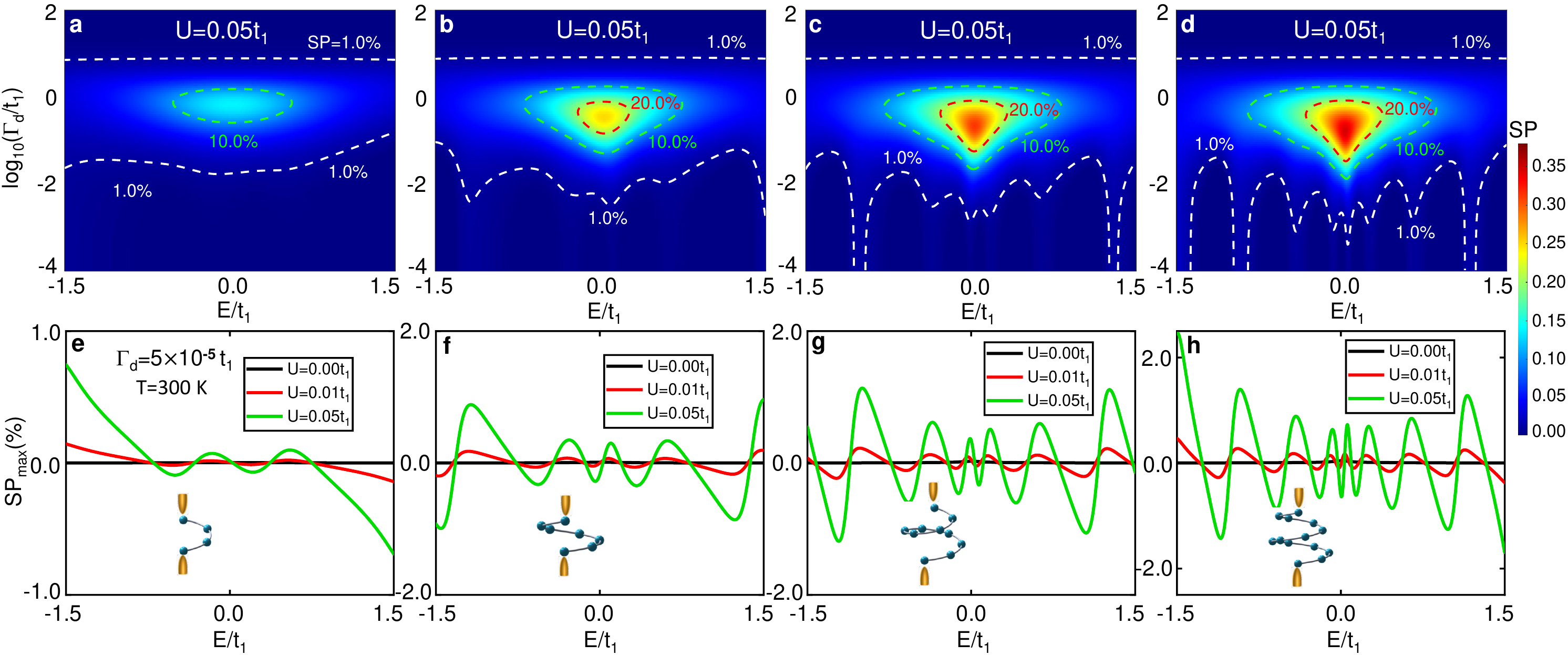}
\caption{Phase diagram of spin polarization and the maximum SP under electron-electron correction. (a)-(d) Phase diagrams of SP for four CSM-based junctions as functions of decoherence strength log$_{10}$($\Gamma_d/t_1$) and onsite energy $E$, calculated at $U$=0.05$t_1$. Contour lines are shown for SP=1.0\%, 10.0\% and 20.0\%. (e)-(h) Maximum SP (SP$_{\mathrm{max}}$) for the same four junctions as a function of $E$, for Hubbard $U=0$, 0.01, and 0.05$t_1$. The decoherence strength and temperature are fixed at $\Gamma_d$=5.0$\times$10$^{-4}t_1$ and 300 K, respectively.}
\label{fig4}
\end{figure*}

While intrinsic SOC fails to explain CISS in irregular helicoids, the present geometric-SOC framework naturally accommodates such geometries. The spatial variation of $\kappa$ for the three typical helix models in Figure 2\textbf{a} is drawn in Figure 2\textbf{c}, which shows that Model B has constant $\kappa$ (uniform SOC), Model A exhibits increasing $\kappa$ with decreasing radius (enhanced SOC), and Model C decreasing $\kappa$ with expanding pitch (suppressed SOC), directly linking local curvature to the geometric SOC strength. Within this framework combined with decoherence, we may trace the evolution of CISS across structural deformation in non-uniform helices. Under moderate decoherence ($\Gamma_d$=0.005$t_1$), the conical helix (Model A, $\Omega$=0.4) exhibits a SP high to 40\% near the Fermi level of $E_F$=0 (Figure 2\textbf{d} with conductance provided in Supporting Information, Section III-D). This value is consistent with the highest experimental SP reported for similar chiral systems\cite{20,21,22,23,24,25,26,27,28}. Note that in the same model, weak decoherence ($\Gamma_d$=0.0001$t_1$) suppresses SP below 1.0\%, while strong decoherence ($\Gamma_d$=100$t_1$) eliminates the CISS signal entirely (Figure 2\textbf{d}). When $\Omega$ decreases and the cone-shape helix transforms into a uniform helical chain (Model B: $\Omega$=$\alpha$=0), the maximum SP falls to about 12\% under the same decoherence (Figure 2\textbf{e}), matching the results obtained using conventional SOC for the same model \cite{37}. Progressing to Model C (increasing $\alpha$) further reduces SP below 1.0\% (Figure 2\textbf{f}), consistent with the weakened geometric SOC due to lower curvature. The continuous variation of SP throughout the structural transition, from cone-shape to uniform and then to progressive-pitch helices, validates the effectiveness and self-consistency of our theoretical approach for irregular helices.\\

\noindent {\fontfamily{SourceSansPro-TLF}\fontseries{b}\selectfont Phase Diagrams of CISS in CSMs.}
To clarify the fundamental reasons behind the presence or absence of CISS in various SCMs and to resolve the experimental contradictions reported for it, we apply this framework to four types of SCMs measured experimentally by Li $et$ $al$. \cite{29}. Based on their geometries and dimensions, we construct 3D tapered helices with half‑turn, one‑turn, and one‑and‑a‑half turns to simulate axial chirality (labeled \textbf{1}S), central chirality (\textbf{2}S, \textbf{3}S), as well as bowl‑shaped SCMs \cite{29}. The corresponding device structures are drawn in the lower right corners of Figures 3\textbf{a}-3\textbf{c}; for comparison, a two-full‑turn helix with varying radius is also considered (Figure 3\textbf{d}). Within the single‑particle picture, we compute SP versus decoherence strength (log$_{10}$($\Gamma_d/t_1$)) and energy $E$ for these four devices, obtaining the phase diagrams in Figures 3\textbf{a}-3\textbf{d}, where contours of SP=1.0\%, 10.0\% and 20.0\% are indicated ($T$=300 K). These diagrams reveal several key common features: (i) observable CISS occurs only in specific parameter regions, that is, significant SP appears only for intermediate decoherence, whereas it becomes negligible under strong coherence or strong decoherence, and also decreases away from the Fermi level. (ii) As molecular size increases, the polarization contours expand, the range of observable SP widens, and the maximum SP rises, indicating that larger molecules favor the observation of CISS, in full agreement with experiments \cite{20,21,22,23,24,25,26,27,28}. (iii) Even molecular fragments that do not form a complete helix (Figure 3\textbf{a}) can produce noticeable CISS, clarifying why some chiral molecules lacking a full helical structure still exhibit this effect. (iv) All phase diagrams are closely linked to molecular handedness; reversing the chirality inverts SP (Supporting Information, Section III-E). 

These calculated results can reasonably account for the conflicting experimental observations in CISS measurements on the same class of CSMs. For instance, in the CISS experiment on CSMs performed by Li $et$ $al.$ using a highly coherent STM‑based break‑junction technique for charge current measurements \cite{29}, the conducting electrons maintain relatively high coherence. This situation corresponds precisely to the regime in the phase diagram where the SP is extremely low (SP$<$1.0\%). Such a low SP would naturally be difficult to detect in electron transport experiments, which is equivalent to saying that the electron transport through the single molecules is too fast to permit electron spin flipping. In contrast, other CISS experiments on CSMs using conventional STM techniques exhibit more pronounced electron decoherence \cite{24,25,26,27,28}. This decoherence environment falls exactly into the high-SP region of the phase diagram, leading to the observation of significant SP. Our framework reconciles three specific discrepancies: (i) the presence vs. absence of CISS in nominally identical molecules \cite{24,29}, (ii) the observation of CISS in non-helical chiral molecules \cite{45_1,45_2,45_3,45_4}, and (iii) the varying SP levels reported for molecules of different sizes.

Although the single-particle framework captures the essential role of decoherence, it does not account for the many-body effects that may further enhance or suppress CISS in CSMs. In what follows, we turn to examine two such effects: electron–electron correction and electron–vibration coupling.\\

\noindent {\fontfamily{SourceSansPro-TLF}\fontseries{b}\selectfont Influences of Electron-Electron Correlations on CISS.}
To extend our geometric-SOC-based framework, we now examine how electron-electron correlations within the single molecules influence the CISS effect across different CSM types. In our calculations, a uniform on-site Coulomb repulsion ($U_i$=$U$) is assumed. We recomputed the phase diagram of SP versus the decoherence strength (log$_{10}$($\Gamma_d/t_1$)) and energy level for the four types of CSMs at $U$=0.05$t_1$; the results are shown in Figures 4\textbf{a}-4\textbf{d} (the phase diagrams for $U$=0.01$t_1$ are provided in Supporting Information, Section V-A). For all CSMs, introducing electron-electron correlation significantly enhances SP, and the three SP contour lines expand notably outward. This behavior is particularly pronounced in the strong-coherence regime. As $U$ further increases, the SP contour lines (e.g., SP=1.0\%) develop distinct resonance peaks. This phenomenon arises because the Coulomb interaction opens a correlation gap in the molecular energy levels, while the Hubbard-type repulsion further splits the spin-up and spin-down levels in nearly degenerate orbitals. In chiral molecules, geometric SOC alone induces spin-dependent transport; the addition of electron–electron correlation amplifies this spin splitting, increases the energy gap between the two spin channels, thereby enhancing spin polarization and giving rise to resonance features. Thus, beyond its role in static level splitting, electron-electron correlation actively amplifies spin selectivity in chiral molecules, particularly in the strong-coherence regime where single-particle CISS would otherwise be absent.

To further highlight the enhancing role of electron–electron correlations in the CISS effect, we selected a region of the phase diagram corresponding to the strong-coherence regime ($\Gamma_d$=5$\times$$10^{-5}t_1$), where CISS signals are typically not detected experimentally \cite{29}, and analyzed how $U$ can enable the observation of CISS signals under such conditions. For the same four types of molecules, the SP as a function of on-site energy $E$ was calculated for Hubbard $U$=0, 0.01$t_1$, and 0.05$t_1$; the results are shown in Figures 4\textbf{e}–4\textbf{h}. When $U$=0 (i.e., within the single-particle framework), no SP occurs. When $U$ takes finite values, the SP as a function of $E$ exhibits clear Coulomb oscillation behavior. As $U$ increases, the oscillation peaks become more pronounced, and the SP correspondingly increases. Notably, larger single molecules display more oscillation peaks and significantly higher polarization values. These oscillation peaks are consistent with the oscillatory behavior of the SP contour lines in the phase diagrams. Thus, in the strong-coherence regime, enhancing electron-electron correlation clearly facilitates the experimental observation of CISS signals, and this enhancement mechanism offers a distinct advantage in larger chiral molecules.\\

\begin{figure*}[t]
\includegraphics[width=2.0\columnwidth]{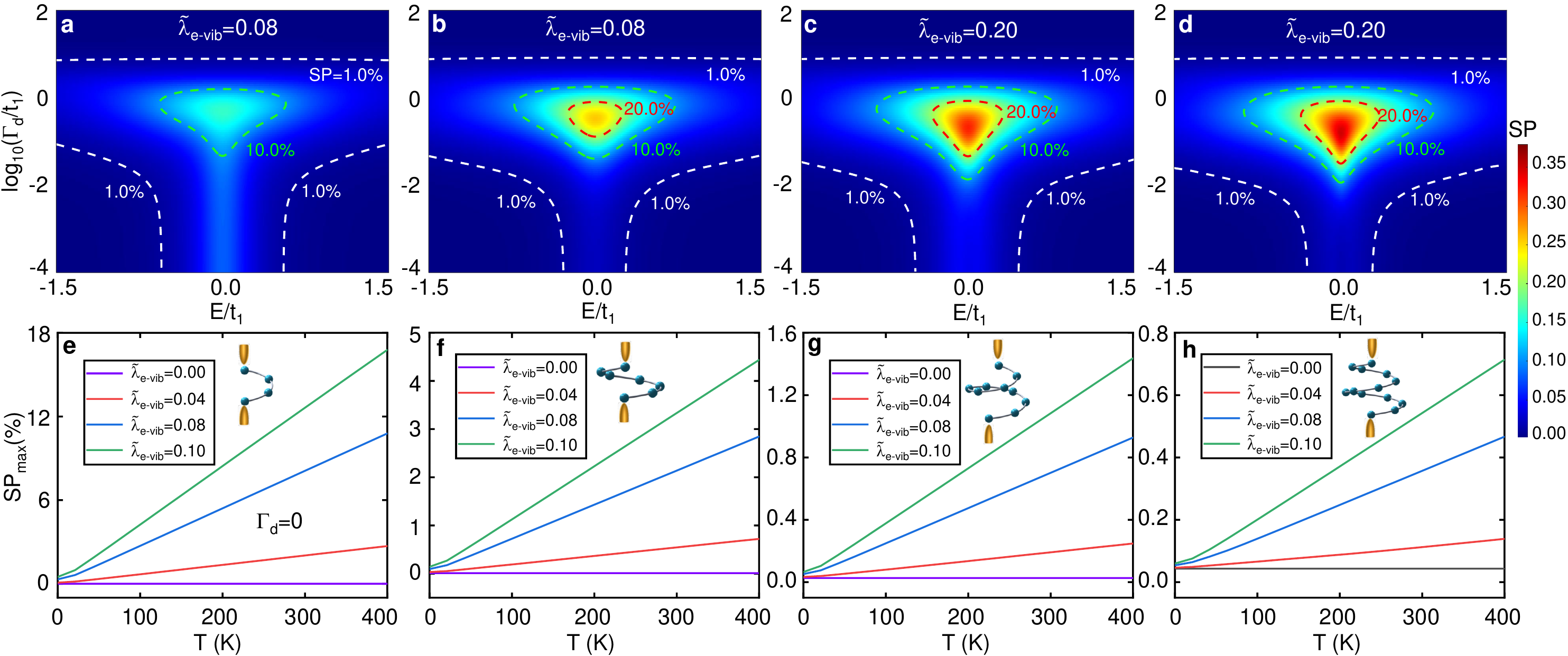}
\caption{Phase diagrams of spin polarization and maximum SP under electron-vibration coupling. (a)-(d) Phase diagrams of SP for four CSM-based junctions as functions of log$_{10}$($\Gamma_d/t_1$) and onsite energy $E$, calculated with electron-vibration coupling ($\tilde\lambda_{\mathrm{e-vib}}$=0.08 or 0.20 $t_1\cdot nm$) and without electron-electron correlation ($U$=0). Contour lines are shown for SP=1.0\%, 10.0\%, and 20.0\%. (e)-(h) Maximum SP (SP$_{\max}$) as a function of temperature for the same four junctions, for $\tilde\lambda_{\mathrm{e-vib}}$=0, 0.04, 0.08, and 0.1 (in unit of $t_1\cdot nm$).}
\label{fig5}
\end{figure*}

\noindent {\fontfamily{SourceSansPro-TLF}\fontseries{b}\selectfont Enhanced CISS by Electron-Vibration Coupling.}
It is well known that another type of many-body interaction in chiral molecules, namely, electron-vibration (e-vib) coupling, also has a significant impact on the CISS effect. Although the theory of e-vib coupling is well established within the conventional SOC framework, how to treat this coupling and construct the corresponding Hamiltonian within the geometric SOC remains an unexplored topic. In real molecules, atoms undergo thermal fluctuations around their equilibrium positions, and this instantaneous position can be expressed as $\vec{r}_{i}$=$\vec{r}^{0}_{i}$+$\vec{Q}_i$ with $\vec{Q}_i$ the displacement vector. The strength of geometric SOC depends on the local curvature $\kappa$, which is determined by the specific positions of atomic arrangement. Within the small-displacement approximation, we expand $\kappa(\vec{r}_{i})$ to linear order around the equilibrium position $\vec{r}^{0}_{i}$ as $\kappa ({\vec{r}_i})\approx\kappa (\vec{r}_i^0) + \nabla \kappa ({\vec{r}_i}){|_{\vec{r}_i^0}} \cdot {\vec{Q}_i}$. Substituting this expansion into the geometric SOC Hamiltonian results in a term that explicitly describes the e-vib coupling:
\begin{align*}
{\mathcal{H}_{e-vib}} = \sum\limits^{N+1}_{i=1} {{\tilde\lambda _{e-vib}}} \left({-\nabla \kappa \left( {{\vec{r}_i}} \right){|_{\vec{r}_i^0}} \cdot {\vec{Q}_i}} \right)\left( {{\vec{N}_i} \times {\vec{p}_i} \cdot {\vec{\sigma}_i}} \right).
\end{align*}
Here $\tilde\lambda _{e-vib}$ denotes the coupling strength, and $\vec{p}_i$, $\vec{N}_i$ and $\vec{\sigma}_i$ are defined as earlier. The full derivation and calculation procedure are provided in Supporting Information, Section IV-B. 

Introducing e-vib coupling significantly alters the SP phase diagrams of the four types of SCMs (Figures 5\textbf{a}-5\textbf{d}, $T$=300 K; results at $T$= 100 and 200 K provided in Supporting Information, Section V-B) compared to the uncoupled case (Figure 3). First, the polarization contours (e.g., SP=1.0\%, 10.0\% and 20.0\%) expand noticeably, indicating an overall stronger polarization, a higher maximum SP, and a substantially broadened energy-decoherence window for observation. Second, the enhancement of SP by this coupling is more pronounced in smaller chiral molecules. These results directly confirm the enhancing effect of e‑vib coupling on CISS. The underlying mechanism stems from the dynamic modulation of spin‑dependent transport channels by vibration modes. Specifically, vibrational degrees of freedom can provide additional energy‑exchange channels for electrons, broadening the distribution of SP along the energy axis, as reflected the widening contour lines in phase diagrams. At the same time, instantaneous vibration fluctuations of molecular geometry, including torsional flexibility and specific bonding motifs, can locally increase curvature, thereby enhancing the instantaneous geometric SOC in the dynamical process. Even in the moderately decoherent region, such fluctuation can still enhance the average polarization. In addition, vibration coupling itself acts as an intrinsic decoherence source, but its synergy with external decoherence helps to stabilize specific spin‑polarized transport channels. This is reflected in the opening of low‑decoherence polarization contours (e.g., 
SP=1.0\%), further supporting a positive correlation between lattice vibrations and decoherence. 

Given that the e-vib coupling can similarly induce decoherence, we adopt it as an alternative description of the decoherence mechanism to re-examine the conditions required for the emergence of CISS. Setting $\Gamma_d$=0 and $U$=0, we calculated the temperature dependence of the maximum SP, SP$_{\max}$, for four CSM-based junctions at e-vib coupling constants $\tilde\lambda_{\mathrm{e-vib}}$=0, 0.04, 0.08, and 0.1 (in unit of $t_1\cdot nm$; with computational methods provided in Supporting Information, Sections II and V-C). The numerical results are shown in Figures 5\textbf{e}-5\textbf{h} and reveal several important trends. First, at $\tilde\lambda_{\mathrm{e-vib}}$=0, none of the CSMs exhibit observable SP; a finite SP appears as soon as $\tilde\lambda_{\mathrm{e-vib}}$ becomes non-zero. This indicates that, under the present conditions, a non-zero e-vib coupling is a necessary condition for the occurrence of CISS. Second, SP$_{\mathrm{max}}$ increases with increasing $\tilde\lambda_{\mathrm{e-vib}}$, demonstrating that within the geometric SOC framework, the e-vib coupling can also serve as an effective mechanism for enhancing CISS. Concurrently, SP$_{\mathrm{max}}$ grows nearly linearly with rising temperature. This trend can account for the experimental observation that sizable spin polarization in CSMs persists at room temperature and may even be further enhanced upon heating. Third, the extent of the CISS enhancement induced by e-vib coupling depends markedly on the molecular geometry. For smaller molecules, the increase in SP$_{\max}$ at a given $\tilde\lambda_{\mathrm{e-vib}}$ is considerably larger than that observed for larger ones. In addition, these results remain stable under moderate on-site energy disorder (Supporting Information, Section V-D). These findings offer theoretical insight into the regulation and enhancement of CISS in single-molecule level, and further corroborate the validity and reliability of the geometric SOC approach for investigating CISS in CSMs.

To better align our theoretical calculations with experimental measurements, we also simulate the spin-polarized tunneling currents driven by a finite bias voltage at room temperature in all four SCM-based junctions (Supporting Information, Section VI), including both electron–vibration coupling and electron–electron interaction under varying decoherence. The resulting SP behavior from the tunneling currents agrees well with the earlier conclusions drawn from spin-polarized conductance, thereby helping to reconcile the apparent discrepancies among CISS experimental results obtained for CSMs under different conditions \cite{27,28,29}. \\

\ACSection{CONCLUSIONS}
\noindent
In summary, we have established a unified theoretical framework for the CISS effect in CSM systems by incorporating geometric SOC and environmental decoherence. This model elucidates the physical mechanisms underlying the CISS effect in CSMs of varying geometries and sizes at finite temperature, both within the single-particle picture and in the presence of many-body correlations. Our results show that the CISS effect exhibits a nonmonotonic dependence on decoherence strength: both the CISS effect and the associated SP are suppressed under strong coherence and strong decoherence, and become significant only within an intermediate decoherence range. This theoretical finding well explains why some experiments on CSMs observe a clear CISS effect with relatively high SP while others do not, thereby clarifying the physical origin of previously contradictory experimental observations on the same CSMs. In particular, our phase diagrams further predict that CISS signals in CSMs are most likely to be observed under moderate decoherence conditions (e.g., $\Gamma_d$$\approx$0.005$t_1$), corresponding to a coherence length of $\sim$10 nm at room temperature. This provides a concrete guideline for designing future single-molecule CISS experiments. 

Furthermore, within the geometric SOC framework, both electron–electron correlation and e–vib coupling can enhance the CISS effect, but their influences differ depending on molecular size. Electron–electron correlation enhances the CISS effect more effectively in larger CSMs, whereas the e-vib coupling is more effective in smaller ones. This work reconciles the conflicting experimental reports on CISS in CSMs and demonstrates that the effect is universal, provided the decoherence strength lies in an intermediate range. The geometric-SOC perspective further shows that CISS is not limited to regular helices but extends to irregular and non-helical chiral structures.\\

\ACSection{ASSOCIATED CONTENT}
\noindent{\bfseries Supporting Information}\\
The Supporting Information is available free of charge at
https://pubs.acs.org/doi/***/***.

\begin{adjustwidth}{2em}{0pt} 
 Geometric SOC, theoretical method to calculate spin-dependent conductance, some results within the single-particle framework, Hamiltonian models and theeretical methods for electron-vibration coupling and electron-electron interaction, influences of many-body correlations, finite temperature and onsite energy disorder on CISS, and spin-dependent currents driven by bias voltages tunneling through all four SCM-based devices.
\end{adjustwidth}
\vspace{0.5\baselineskip}
\ACSection{AUTHOR INFORMATION}
\noindent\hspace*{-0.3em}\ACSSISubtitle{Corresponding Author}
\begin{adjustwidth}{1.5em}{0pt} 
\textbf{Hua-Hua Fu} -- 
\textit{School of Physics and Wuhan National High Magnetic Field Center,
Huazhong University of Science and Technology, Wuhan 430074, People's Republic of China;
Institute for Quantum Science and Engineering, Huazhong University of Science and Technology, Wuhan 430074, People's Republic of China;} %
\orcidicon~\href{https://orcid.org/0000-0003-3920-6324}{orcid.org/0000-0003-3920-6324};\,
Email: \href{mailto:hhfu@hust.edu.cn}{hhfu@hust.edu.cn}.
\end{adjustwidth}
\vspace{0.5\baselineskip}

\noindent\hspace*{-0.3em}\ACSSISubtitle{Authors}
\begin{adjustwidth}{1.5em}{0pt}

\textbf{Shu-Zheng Zhou} --
\textit{School of Physics and Wuhan National High Magnetic Field Center,
Huazhong University of Science and Technology, Wuhan 430074, People's Republic of China.} %

\noindent
\textbf{Xi Sun} --
\textit{School of Physics and Wuhan National High Magnetic Field Center,
Huazhong University of Science and Technology, Wuhan 430074, People's Republic of China.} %

\noindent
\textbf{Kai-Yuan Zhang} --
\textit{School of Physics and Wuhan National High Magnetic Field Center,
Huazhong University of Science and Technology, Wuhan 430074, People's Republic of China.} %

\end{adjustwidth}

\vspace{0.5\baselineskip}

\noindent\hspace*{-0.5em}\ACSSISubtitle{Notes}\\
The authors declare no competing financial interest.\\

\ACSection{ACKNOWLEDGEMENTS}
\noindent This work is supported by authors' personal resource.\\

\noindent{\bfseries REFERENCES}

\clearpage
\onecolumngrid





\begin{thebibliography}{99}
\newcommand{\DOI}[1]{doi: \href{https://doi.org/#1}{#1}}

\bibitem{1} Naaman, R.; Paltiel, Y.; Waldeck, D. H. Chiral molecules and the electron spin. \textit{Nat. Rev. Chem.} \textbf{2019}, \textit{3}, 250-260.

\bibitem{2} Yang, S.-H.; Naaman, R.; Paltiel, Y.; Parkin, S. S. P. Chiral spintronics. \textit{Nat. Rev. Phys.} \textbf{2021}, \textit{3}, 328.

\bibitem{3} Aiello, C. D.; Abendroth, J. M.; Abbas, M.; Afanasev, A.; Agarwal, S.; Banerjee, A. S.; Beratan, D. N.; Belling, J. N.; Berche, B.; Botana, A.; et al. A Chirality-Based Quantum Leap. \textit{ACS Nano} \textbf{2022}, \textit{16}, 4989-5035.

\bibitem{4} Bloom, B. P.; Chen, Z.; Lu, H.; Waldeck, D. H. A. Chemical perspective on the chiral induced spin selectivity effect, \textit{Nat. Sci. Rev.} \textbf{2024}, \textit{11}, nwae212.

\bibitem{5} Bloom, B. P.; Paltiel, Y.; Naaman, R.; Waldeck, D. H. Chiral Induced Spin Selectivity. \textit{Chem. Rev.} \textbf{2024}, \textit{124}, 1950-1991.

\bibitem{6} Ray, K.; Ananthavel, S. P.; Waldeck, D. H.; Naaman, R. Asymmetric Scattering of Polarized Electrons by Organized Organic Films of Chiral Molecules, \textit{Science} \textbf{1999}, \textit{283}, 814.

\bibitem{7} G\"{o}hler, B.; Hamelbeck, V.; Markus, T. Z.; Kettner, M.; Hanne, G. F.; Vager, Z.; Naaman, R.; Zacharias, H. Spin Selectivity in Electron Transmission Through Self-Assembled Monolayers of Double-Stranded DNA, \textit{Science} \textbf{2011}, \textit{331}, 894.

\bibitem{8} Inui, A.; Aoki, R.; Nishiue, Y.; Shiota, K.; Kousaka, Y.; Shishido, H.; Hirobe, D.; Suda, M.; Ohe, J.; Kishine, J.; Yamamoto, H. M.; Togawa, Y. Chirality-Induced Spin-Polarized State of a Chiral Crystal CrNb$_3$S$_6$. \textit{Phys. Rev. Lett.} \textbf{2020}, \textit{124}, 166602.

\bibitem{9} Shiota, K.; Inui, A.; Hosaka, Y.; Amano, R.; \={O}nuki, Y.; Hedo, M.; Nakama, T.; Hirobe, D.; Ohe, J.; Kishine, J.; Yamamoto, H. M.; Shishido, H.; Togawa, Y. Chirality-Induced Spin-Polarized over Macroscopic Distances in Chiral Disilicide Crystals. \textit{Phys. Rev. Lett.} \textbf{2021}, \textit{127}, 126602.


\bibitem{10} Rodr\'{\i}guez, R.; Naranjo, C.; Kumar, A.; Matozzo, P.; Das, T. K.; Zhu, Q.; Vanthuyne, N.; G\'{o}mez, R.; Naaman, R.; S\'{a}nchez, L.; Crassous, J. Mutual Monomer Orientation To Bias the Supramolecular Polymerization of [6]Helicenes and the Resulting Circularly Polarized Light and Spin Filtering Properties. \textit{J. Am. Chem. Soc.} \textbf{2022}, \textit{144}, 7709-7719.

\bibitem{11} Giaconi, N.; Poggini, L.; Lupi, M.; Briganti, M.; Kumar, A.; Das, T. K.; Sorrentino, A. L.; Viglianisi, C.; Menichetti, S.; Naaman, R.; Sessoli, R.; Mannini, M. Efficient Spin-Selective Electron Transport at Low Voltages of Thia-Bridged Triarylamine Hetero[4]helicenes Chemisorbed Monolayer. \textit{ACS Nano} \textbf{2023}, \textit{17}, 151189-15198.

\bibitem{12} Jiang, H.; \v{C}avlovi\'{c}, D.; Jiang, Q.; Ng, F.; Bao, S. T.; Telford, E. J.; Steigerwald, M. L.; Roy, X.; Nuckolls, C.; McNeill, J. M. Spin Filtering with Surface-Active Helicene- and Twistacene-Based Perylene Diimides. \textit{J. Am. Chem. Soc.} \textbf{2025}, \textit{147}, 12982-12988.

\bibitem{13} Singh, A-K.; Martin, K.; Talamo, M. M.; Houssin, A.; Vanthuyne, N.; Avarvari, N.; Tal, O. Single-molecule junctions map the interplay between electrons and chirality. \textit{Nat. Commun.} \textbf{2025}, \textit{16}, 1759.


\bibitem{14} Lu, H.; Wang, J.; Xiao, C.; Pan, X.; Chen, X.; Brunecky, R.; Berry, J. J.; Zhu, K.; Beard, M. C.; Vardeny, Z. V. Spin-dependent charge transport through 2D chiral hybrid lead-iodide perovskites. \textit{Sci. Adv.} \textbf{2019}, \textit{5}, 0571.

\bibitem{15} Kim, Y.-H.; Zhai, Y.; Lu, H.; Pan, X.; Xiao, C.; Gaulding, E. A.; Harvey, S. P.; Berry, J. J.; Vardeny, Z. V.; Luther, J. M.; Beard, M. C. Chiral-induced spin selectivity enables a room-temperature spin light-emitting diode. \textit{Science} \textbf{2021}, \textit{371}, 1129-1133.

\bibitem{16} Qian, Q.; Ren, H.; Zhou, J.; Wan, Z.; Zhou, J.; Yan, X.; Cai, J.; Wang, P.; Li, B.; Sofer, Z.; Li, B.; Duan, X.; Pan, X.; Huang, Y.; Duan, X. Chiral molecular intercalation superlattices. \textit{Nature (London)} \textbf{2022}, \textit{606}, 902.

\bibitem{17} Xu, Y.; Mi, W. Chiral-induced spin selectivity in biomolecules, hybrid organic–inorganic perovskites and inorganic materials: a comprehensive review on recent progress. \textit{Mater. Horiz.} \textbf{2023}, \textit{10}, 1924.

\bibitem{18} Lv, J.; Gao, X.; Han, B.; Zhu, Y.; Hou, K.; Tang, Z. Self-assembled inorganic chiral superstructures. \textit{Nat. Rev. Chem.} \textbf{2022}, \textit{6}, 125-145.

\bibitem{19} Wang, C.; Liang, Z.-R.; Chen, X.-F.; Guo, A.-M.; Ji, G.; Sun, Q.-F.; Yan, Y. Transverse Spin Selectivity in Helical Nanofibers Prepared without Any Chiral Molecule. \textit{Phys. Rev. Lett.} \textbf{2024}, \textit{133}, 108001.


\bibitem{20} Eckvahl, H. J.; Tcyrulnikov, N. A.; Chiesa, A.; Bradley, J. M.; Young, R. M.; Carretta, S.; Krzyaniak, M. D.; Wasielewski, M. R. Direct observation of chirality-induced spin selectivity in electron donor-acceptor molecules. \textit{Science} \textbf{2023}, \textit{382}, 197-201.

\bibitem{21} M\"{o}llers, P. V.; Urban, A. J.; Feyter, S. D.; Yamamoto, H. M.; Zacharias, H. Probing the Roles of Temperature and Cooperative Effects in Chirality-Induced Spin Selectivity: Photoelectron Spin Polarization in Helical Tetrapyrroles. \textit{J. Phys. Chem. Lett.} \textbf{2024}, \textit{15}, 9620-9629.

\bibitem{22} Niu, W.; Fang, C.; Tang, L.; Unsal, E.; Fu, Y.; Deka, J.; Liu, F.; Popov, A. A.; Wu, F.; Shi, H.; Komber, H.; Dianat, Z.; Gutierrez, R.; Ma, J.; Sang, Y.; Cuniberti, G.; Parkin, S. S. P. Lateral $\pi$-extended helical nanographenes with large spin polarization. \textit{Chem. Sci.} \textbf{2025}, \textit{16}, 21446-21453.

\bibitem{23} Safari, M. R.; Matthes, F.; Schneider, C. M.; Ernst, K.-H.; B\"{u}rgler, D. Spin-Selective Electron Transport Through Single Chiral Molecules. \textit{Small} \textbf{2024}, \textit{20}, 2308233.

\bibitem{24} Eckvahl, H. J.; Copley, G.; Young, R. M.; Krzyaniak, M. D.; Wasielewski, M. R. Detecting Chirality-Induced Spin Selectivity in Randomly Oriented Radical Pairs Photogenerated by Hole Transfer. \textit{J. Am. Chem. Soc.} \textbf{2024}, \textit{146}, 24125-24132.

\bibitem{25} Artemyev, A. N.; Tomar, R.; Trabert, D.; Kargin, D.; Kutscher, E.; Sch\"{o}ffler, M. S.; Schmidt, L. P. H.; Pietschnig, R.; Jahnke, T.; Kunitski, M.; Eckart, S.; D\"{o}rner, R.; Demekhin, P. V. Photoelectron Circular Dichroism in the Spin-Polarized Spectra of Chiral Molecules. \textit{Phys. Rev. Lett.} \textbf{2024}, \textit{132}, 123202.

\bibitem{26} Zhang, D.-Y.; Sang, Y.; Das, T. K.; Guan, Z.; Zhong, N.; Duan, C.-G.; Wang, W.; Fransson, J.; Naaman, R.; Yang, H.-B. Highly Conductive Topologically Chiral Molecular Knots as Efficient Spin Filters. \textit{J. Am. Chem. Soc.} \textbf{2025}, \textit{145}, 26791-26798.

\bibitem{27} Labella, J.; Gupta, A.; Kumar, A.; L\'{o}pez-Serrano, E.; Bhowmick, D. K.; Naaman, R.; Torres, T. Spin filtering in self-assembled bowl-shaped aromatics. \textit{Chem. Sci.} \textbf{2025}, \textit{16}, 8783-8787.

\bibitem{28} Wang, Y.; Zhang, Y.; Wang, Y.-Y.; Yan, Q. Highly Conductive Chiral Organic Cages and Their Helical Assemblies Enable Efficient Spin Filtering. \textit{J. Am. Chem. Soc.} \textbf{2025}, \textit{147}, 8751-8759.


\bibitem{29} Li, L.; Shi, W.; Mahajan, A.; Zhang, J.; G\'{o}mez-G\'{o}mez, M.; Labella, J.; Louie, S.; Torres, T.; Barlow, S.; Marder, S. R.; Reichman, D. R.; Venkataraman, L. Too Fast for Spin Flipping: Absence of Chirality-Induced Spin Selectivity in Coherent Electron Transport through Single-Molecules Junctions. \textit{J. Am. Chem. Soc.} \textbf{2025}, \textit{147}, 25043.

\bibitem{29_1} Evers, F.; Koryt\'{a}r, R.; Tewari, S.; van Ruitenbeek, J. M. Advances and challenge in single-molecule electron transport. \textit{Rev. Mod. Phys.} \textbf{2020}, \textit{92}, 035001.

\bibitem{30} Dalum, S.; Hedeg\aa rd, P. Theory of chiral induced spin selectivity. \textit{Nano Lett.} \textbf{2019}, \textit{19}, 5253.

\bibitem{31} Li, X.; Nan, J.; Pan, X. Chiral Induced Spin Selectivity As a Spontaneous Intertwined Order. \textit{Phys. Rev. Lett.} \textbf{2020}, \textit{125}, 263002.

\bibitem{32} Liu, Y.; Xiao, J.; Koo, J.; Yan, B. Chirality-driven topological electronic structure of DNA-like materials. \textit{Nat. Mater.} \textbf{2021}, \textit{20}, 638-644.

\bibitem{33} Alwan, S.; Dubi, Y. Spinterface origin for the chirality-induced spin-selectivity effect. \textit{J. Am. Chem. Soc.} \textbf{2021}, \textit{143}, 14235.

\bibitem{34} Evers, F.; Aharony, A.; Bar-Gill, N.; Entin-Wohlman, O.; Hedeg{\aa}rd, P.; Hod, O.; Jelinek, P.; Kamieniarz, G.; Lemeshko, M.; Michaeli, K.; et. al. Theory of Chirality Induced Spin Selectivity: Progress and Challenges. \textit{Adv. Mater.} \textbf{2022}, \textit{34}, 2106629.

\bibitem{35} Wolf, Y.; Liu, Y.; Xiao, J.; Park, N.; Yan, B. Unusual Spin Polarization in the Chirality-Induced Spin Selectivity. \textit{ACS Nano} \textbf{2022}, \textit{16}, 18601.

\bibitem{36} Dednam, W.; Garc{\'\i}a-Bl{\'a}zquez, M. A.; Zotti, L. A.; Lombardi, E. B.; Sabater, C.; Pakdel, S.; Palacios, J. J. A Group-Theoretic Approach to the Origin of Chirality-Induced Spin-Selectivity in Nonmagnetic Molecular Junctions. \textit{ACS Nano} \textbf{2023}, \textit{17}, 6452.


\bibitem{37} Guo, A.-M.; Sun, Q.-F. Spin-dependent electron transport in protein-like single-helical molecules. \textit{Proc. Natl. Acad. Sci. U.S.A.} \textbf{2014}, \textit{111}, 11658-11662.

\bibitem{38} Zhang, L.; Hao, Y.; Qin, W.; Xie, S.; Qu, F. Chiral-induced spin selectivity: A polaron transport model. \textit{Phys. Rev. B} \textbf{2020}, \textit{102}, 214303.

\bibitem{39} Volosniev, A. G.; Alpern, H.; Paltiel, Y.; Millo, O.; Lemeshko, M.; Ghazaryan, A. Interplay between friction and spin-orbit coupling as a source of spin polarization. \textit{Phys. Rev. B} \textbf{2021}, \textit{104}, 024430.

\bibitem{40} Matityahu, S.; Utsumi, Y.; Aharony, A.; Entin-Wohlman, O.; Balseiro, C. A. Spin-dependent transport through a chiral molecule in the presence of spin-orbit interaction and nonunitary effects. \textit{Phys. Rev. B} \textbf{2016}, \textit{93}, 075407.

\bibitem{41} Adhikari, Y.; Liu, T.; Wang, H.; Hua, Z.; Liu, H.; Lochner, E.; Schlottmann, P.; Yan, B.; Zhao, J.; Xiong, P. Interplay of structural chirality, electron spin and topological orbital in chiral molecular spin valves. \textit{Nat. Commun.} \textbf{2023}, \textit{14}, 5163.

\bibitem{42} Aharony, A.; Entin-Wohlman, O. Spin–orbit interactions, time-reversal symmetry, and spin selection. \textit{J. Chem. Phys.} \textbf{2025}, \textit{162}, 154103.

\bibitem{43} Chen, S.; Wu, R.; Fu, H.-H. Persistent Chirality-Induced Spin-Selectivity Effect in Circular Helix Molecules. \textit{Nano Lett.} \textbf{2024}, \textit{24}, 6210-6217.

\bibitem{44} Chen, S.; Fu, H.-H. Chirality-Induced Majorana Zero Modes and Majorana Polarization. \textit{ACS Nano} \textbf{2024}, \textit{18}, 34126-34133.

\bibitem{45} Sun, X.; Zhang, K.-Y.; Zhou, S.-Z.; Fu, H.-H. Robust chirality-induced spin selectivity in topologically chiral molecular knots. \textit{Nat. Commun.} \textbf{2026}, \textit{17}, 1231.

\bibitem{45_01} Alwan, S.; Dubi, Y. Spinterface Origin for the Chirality-Induced Spin-Selectivity Effect. \textit{J. Am. Chem. Soc.} \textbf{2021}, \textit{143}, 14235-14241.


\bibitem{45_1} Wang, Y.; Zhang, Y.; Wang, Y.-Y.; Yan, Q. Highly Conductive Chiral Organic Cages and Their Helical Assemblies Enable Efficient Spin Filtering. \textit{J. Am. Chem. Soc.} \textbf{2025}, \textit{147}, 8751-8759.

\bibitem{45_2} Labella, J.; Osterloh, W. R.; Kuo, K.; Tsutsui, Y.; Tanaka, T.; Seki, S. A Stable, NH-Containing Chiral Nanographene: An Electroactive N-Doped $\pi$-System for Chiroptics and Spin-Selective Transport. \textit{J. Am. Chem. Soc.} \textbf{2026}, \textit{148}, 9670-9679.

\bibitem{45_3} Jiang, C.; Jin, C.; Lv, Y.; Han, X.; Li, Z.; Liu, Z.; Yan, Q.; Liu, Y.; Dai, S.; Li, D.; Cui, Y. Topological-Controlled Chirality and Spin Selectivity in Two-Dimensional Covalent Organic Frameworks. \textit{J. Am. Chem. Soc.} \textbf{2026}, \textit{148}, 2551-2562.

\bibitem{45_4} Rana, S.; Remigo, M.; Geetha, L. A.; Strutynski, K.; Volpi, M.; John, S.; Baczewski, L. T.; Paltiel, Y.; Resel, R.; Melle-Franco, M.; Mali, K. S.; Geerts, Y. H.; Feyter, S. D. Chirality-Induced Spin Selectivity in Two-Dimensional Self-Assembled Molecular Networks. \textit{J. Am. Chem. Soc.} \textbf{2025}, \textit{147}, 42426-42432.

\bibitem{46} Shitade, A.; Minamitani, E. Geometric spin–orbit coupling and chirality-induced spin selectivity. \textit{New J. Phys.} \textbf{2020}, \textit{22}, 113023.

\bibitem{47} Yu, Z.-G. Chirality-Induced Spin-Orbit Coupling, Spin Transport, and Natural Optical Activity in Hybrid Organic-Inorganic Perovskites. \textit{J. Phys. Chem. Lett.} \textbf{2020}, \textit{11}, 8638.

\bibitem{48} Yu, Z.-G. Spin-Charge Conversion in Chiral Polymers with Hopping Conduction. \textit{J. Phys. Chem. Lett.} \textbf{2024}, \textit{15}, 7770-7774.

\bibitem{48_1} Kuemmeth, F.; Ilani, S.; Ralph, D. C.; McEuen, P. L. Coupling of spin and orbital motion of electrons in carbon nanotubes. \textit{Nature} \textbf{2008}, \textit{452}, 448-452.

\bibitem{48_2} Naaman, R.; Paltiel, Y. Why Is the Mechanism Underlying the Chiral-Induced Selectivity Effect Still Challenging? \textit{Adv. Mater.} \textbf{2026}, \textit{0}, e23675.

\bibitem{49} Cheng, S.; Zhang, K.-Y.; Sun, X.; Zhou, S. Z.; Fu, H.-H. Synergy and Competition of Dual Chirality in the Chirality-Induced Spin Selectivity of Supramolecular Helices. \textit{J. Am. Chem. Soc.} (Accepted in Publication).

\bibitem{49_1} Xing, Y.; Sun, Q.-F.; Wang, J. Influence of dephasing on the quantum Hall effect and the spin Hall effect. \textit{Phys. Rev. B} \textbf{2008}, \textit{77}, 115346.

\bibitem{49_2} Jiang, H.; Cheng, S.; Sun, Q.-F.; Xie, X. C. Topological Insulator: A New Quantized Spin Hall Resistance Robust to Dephasing. \textit{Phys. Rev. Lett.} \textbf{2009}, \textit{103}, 036803.

\bibitem{50} Fu, Y.; Dudley, S. C. Quantum inductance within linear response theory. \textit{Phys. Rev. Lett.} \textbf{1993}, \textit{70}, 65-68.

\bibitem{51} Nemnes, G. A.; Wulf, U.; Racec, P. N. Nano-transistors in the Laudauer-Buttiker formalism. \textit{J. Appl. Phys.} \textbf{2004}, \textit{96}, 596-604.

\bibitem{52} Guo, A.-M.; Sun, Q.-F. Sequence-dependent spin-selective tunneling along double-stranded DNA. \textit{Phys. Rev. B} \textbf{2012}, \textit{86}, 115441.

\bibitem{53} Du, G.-F.; Fu, H.-H.; Wu, R. Vibration-enhanced spin-selective transport of electrons in the DNA double helix. \textit{Phys. Rev. B} \textbf{2020}, \textit{102}, 035431.

\end{thebibliography}
\end{document}